%% file: Paper.tex
\documentclass[12pt]{article}

\input{econfmacros}
\input econfmacros.tex
\def\Title#1{\begin{center} {\Large {\bf #1} } \end{center}}

\usepackage{jheppub}
\usepackage{atlasphysics} 
\usepackage{upgreek}
 \usepackage{lineno}
\input{keys}

\begin{document}
\Title{Measurement of \DGs\ and the $CP$-violating weak phase \phis\  in the decay \Bstbold\ by ATLAS}

\bigskip


\begin{raggedright}  

{\it Maria Smizanska, on behalf of the ATLAS Collaboration, \index{Smizanska, M.} \\ Lancaster University, UK.
}
\end{raggedright}

\section{Introduction}
 New phenomena beyond the predictions of the Standard Model (SM) may alter $CP$ violation in $B$-decays. In the decay \Bst \  the $CP$-violating phase \phis\ is  the weak phase difference between the \BsaBs mixing amplitude and the $b \rightarrow c \overline{c} s$ decay amplitude.   SM  predicts a small value of  $\phis \simeq -2 \beta_{s} = -0.0368 \pm 0.0018 $ rad \cite{PHISM}. 
Another quantity involved in \BsaBs mixing is the width difference $\DGs = \GL - \GH$ of heavy (\BH) and light (\BL) eigenstates.  Physics beyond the SM is not expected to affect \DGs\ significantly~\cite{Nierste}. Extracting \DGs\ from data is nevertheless useful as it allows theoretical predictions to be tested~\cite{Nierste}. 
The presented analysis uses data collected by the ATLAS experiment in  $pp$ collisions at $\sqrt{s} = 7 \TeV$ in 2011, corresponding to an integrated luminosity of  $\sim 4.9~ \ifb$. Measurements of \phis, the average decay width $\Gs = (\Gamma_L + \Gamma_H) / 2$ and the  value  of \DGs,   were performed with no flavour tagging to distinguish between   the initial \Bs\ and \aBsn\ states. The $CP$ states were separated statistically through the time-dependence of the decay and angular correlations amongst the final-state particles.

\section{Candidate selection } \label{sec:sample}
ATLAS  is a multipurpose detector described in details in \cite{ATLAS}. The  tracking and muon systems are of particular importance in the reconstruction of $B$~mesons. The inner tracking detector  consists of a silicon pixel detector, a silicon microstrip detector  and a transition radiation tracker; all surrounded by a superconducting solenoid providing a 2\,T axial field. The muon spectrometer consists of three  superconducting toroids, a system of tracking chambers, and detectors for triggering.  The triggers used  for this analysis are based on identification of a \Jpsitomm decay, with either a $4\;\mathrm{GeV}$ transverse momentum (\pT) threshold for each muon or an asymmetric configuration that applies a higher \pT\ threshold ($4-10\GeV$) to one of the muons.
The pairs of muon tracks  refitted to a common vertex are accepted if the fit results in \cutChiJpsi \ and the invariant mass  falls in the range \cutMassJpsiBB (when both muons have $|\eta| <  1.05$), or  \cutMassJpsiEE (both muons within $1.05  < |\eta| <  2.5 $)  or  \cutMassJpsiEB \  otherwise. Candidates  for  \Bsto\ are sought by fitting the four tracks to a common vertex resulting in  \cutChiBs,  while  the invariant mass of the two muons was fixed to the    \Jpsipa mass~\cite{PDG}. In total $131\mathrm{k}$ \Bs\ candidates are collected within a mass range of   \phiKK\  (1.0085  - 1.0305)~GeV and    $5.15 < m(\Bs) < 5.65 \GeV$, figure \ref{fig:simulFittime}.
For each \Bs\  candidate the proper decay time $t$  ('decay time' henceforth)  is determined by the expression:
 $  t =  L_{xy}\ M_{{}_{B}}   /   ( c \ p_{ \mathrm{T}_{B}  } )$, 
where $p_{ \mathrm{T}_{B}  }$ is the transverse momentum of the \Bs\  candidate and  $M_{{}_{B}}$ is the mass of the \Bs\ meson (\BspdgNoerr) \cite{PDG}. $L_{xy}$ is the displacement in the transverse plane of the  \Bs\  decay vertex with respect to the primary vertex (PV)  projected onto the direction of $p_{ \mathrm{T}_{B}  }$. The position of the PV is refitted following the removal of the tracks used to reconstruct the \Bs\  candidate. For the selected events the average number of pileup interactions is 5.6, necessitating  a choice of the best candidate for the PV at which the \Bs\   is  produced.  The variable used  is a three-dimensional impact parameter $d_0$,  calculated as the distance between the line extrapolated from the \Bs\  vertex in the~direction of the \Bs\ momentum, and  each PV candidate. The chosen PV is the one with the smallest~$d_0$. 

\section{Analysis and results}

The \Bst \ decay involves three angular momentum states of the \JpsiPhi \ system, combined into three polarization amplitudes, longitudinal polarization ($A_0$), and transverse polarization with the linear polarization vectors of the vector mesons parallel ($A_{\parallel}$) or perpendicular ($A_{\perp}$) to each other. The first two states are $CP$-even, while the last state is $CP$-odd. 
Another $CP$-odd state can be produced by a non-resonant  $K^+K^-$ pair or by the decay of the spin-0 $f_0$  meson, resulting in another independent amplitude, the S-wave $A_S$. The time evolution of the four decay amplitudes along with six interference terms is fitted simultaneously  with  the   angular distributions of the  cascade decay \Bsto. Three independent angles $\Omega = (\theta_T,\psi_T,\varphi_T)$, are defined in the transversity basis  \cite{CDFAngles}.    The \Bst \ decay probability  is a function  of the following physics parameters of interest,  \DGs, \phis, \Gs, the amplitudes  $A_{0}$, $A_{\parallel}$, $A_{S}(0)$ and of the related strong phases defined as $\delta_{||} $=   arg($A_{\parallel}, A_{0}$), $\delta_{\perp}$ = arg($A_{\perp}, A_{0}$) and $\delta_{S}$ = arg($A_{S}, A_{0}$). For  untagged analysis all  terms involving the mass splitting $\Delta m_s$  in the time-dependent amplitudes cancel out  \cite{ATLASBs}.  In addition, the time-dependent amplitudes depending on $\delta_{\perp}$ are multiplied by a small value of $\sin\phis$, hence the untagged analysis is not sensitive to $\delta_{\perp}$. A Gaussian constraint to the external data, $\delta_{\perp} = (2.95 \pm 0.39)$~rad \cite{LHCb:2011aa} is therefore applied.

An unbinned maximum likelihood fit  uses information of the reconstructed \Bs\ candidate mass $m$, decay time $t$, their  uncertainties $\sigma_m$ and $\sigma_t$,  and the angles  $\Omega = (\theta_T,\psi_T,\varphi_T)$. In total 26 parameters are determined, including the eight physics parameters of interest mentioned  above,  while  the other  quantities describe the $J/\psi \phi$ mass distribution,  the decay time and the  angular distributions of the background. The single-event likelihood has the form:
\begin{eqnarray}
{\cal L}  \propto w_i \cdot   f_{\textrm{s}} \cdot   {\cal F}_{\textrm{s}i} +  f_{\textrm{s}} \cdot f_{B^0} \cdot {\cal F}_{B_i^0} 
 + ( 1 -  f_{\textrm{s}} \cdot (1+ f_{B^0 } )    ) \cdot {\cal F}_{\textrm{bkg}i}  \   \nonumber
\end{eqnarray}
where the index $i$ is used for the variables specific  for each single-event,   $f_\textrm{s}$ is the fraction of signal candidates, ${ \cal F}_{\textrm{s} i}$ and  ${\cal F}_{\textrm{bkg}i }$  are probability density functions (PDF) modelling the signal and background.  The  backgrounds \Bdt\ and $\Bd\rightarrow J/\psi K^{+}\pi^{-}$   are parametrized separately by ${\cal F}_{B_i^0} $ with  $f_{B^0}$  being the fraction of  this  background events. The weighting factor $w_i$ accounts for a small  decay-time dependency of the acceptance, 
related to a limited resolution in the on-line track reconstruction, details are given in  \cite{ATLASBs}.
The  PDF describing  the signal, ${\cal F}_{\textrm{s}}$,   has the form:
\begin{eqnarray} 
  { \cal F}_{\textrm{s} i} &=&  P_\textrm{s}(m_{i}|\sigma_{m_{i}}) \cdot P_\textrm{s}(\sigma_{m_{i}})\cdot P_\textrm{s}(\Omega_{i}, t_{i}|\sigma_{t_{i}}) \cdot P_\textrm{s}(\sigma_{t_{i}}) \cdot A(\Omega_i,{\it p_{\mathrm{T}i}})  \cdot P_\textrm{s}({\it p_{\mathrm{T}i}}) \nonumber
 \end{eqnarray}
The signal mass density $P_\textrm{s}(m_{i}|\sigma_{m_{i}})$ is  modelled  as a Delta function smeared by a Gaussian with a per-candidate mass resolution $\sigma_{m_{i}}$. Similarly, each of ten terms of the signal  time and angular dependence, $P_\textrm{s}(\Omega_{i}, t_{i}|\sigma_{t_{i}})$, is convoluted  with a Gaussian with a per-candidate resolution  $\sigma_t$. The angular sculpting of the detector and kinematic cuts on the angular distributions is included in the likelihood function through $A(\Omega_i, {\it p_{\mathrm{T}i}})$. This is calculated using a four-dimensional binned acceptance method, applying an event-by-event efficiency according to the transversity angles ($\theta_T,\psi_T,\varphi_T$) and the $p_{ \mathrm{T}_{B}  }$. The acceptance was calculated from the \Bst\ MC events. The background PDF has the following composition:
\begin{eqnarray} \label{eq:likelihoodBkg}
  {\cal F}_{\textrm{bkg}i }  =   P_\textrm{b}(m_{i})\cdot P_\textrm{b}(\sigma_{m_{i}}) \cdot P_\textrm{b}(t_{i}|\sigma_{t_{i}})                                                             \cdot  P_\textrm{b}(\theta_T)\cdot P_\textrm{b}(\varphi_T) \cdot P_\textrm{b}(\psi_T)\cdot P_\textrm{b}(\sigma_{t_{i}})  \cdot P_\textrm{b}({\it p_{\mathrm{T}i}}) \nonumber
\end{eqnarray}
The decay time function $P_\textrm{b}(t_{i}|\sigma_{t_{i}})$ is parameterised as a Delta function, two positive exponentials and a negative exponential. These functions are smeared with the same resolution function as the signal decay time-dependence. The prompt peak models the combinatorial background events, which are expected to have reconstructed lifetime distributed around zero. The two positive exponentials represent a  fraction of longer-lived backgrounds with non-prompt $J/\psi$, combined with hadrons from the primary vertex or  from a  $ B/D$  hadron in the same event.  The negative exponential takes into account events with poor vertex resolution.  
The shape of the background angular distributions, $P_\textrm{b}(\theta_T)$, $P_\textrm{b}(\varphi_T)$, and $P_\textrm{b}(\psi_T)$  arising primarily from detector and kinematic sculpting are  described by the  empirical functions with nuisance parameters determined in the fit. 
 The correlations between the background angular shapes are neglected, but a systematic error arising from this simplification was evaluated. The background mass model, $P_\textrm{b}(m)$ is a linear function.
 Mis-reconstructed \Bdt\ and $\Bd\rightarrow J/\psi K^{+}\pi^{-}$ (non-resonant) decays, are parametrized separately. 
The fractions of these components are fixed in the likelihood fit to values $(6.5 \pm 2.4)\%$ and $(4.5 \pm 2.8) \% $, calculated using MC events. Mass and angles have fixed shapes determined from~the~MC studies. The  decay time is described  by an exponential smeared with per-candidate Gaussian errors. Finally, the terms  $ P_\textrm{s,b}(\sigma_{m_{i}})$,  $P_\textrm{s,b}(\sigma_{t_{i}}) $ and   $P_\textrm{s,b}({\it p_{\mathrm{T}i}}) $  are introduced to account for differences between signal and background per-candidate mass and decay time uncertainties and values of transverse momenta, details  are given in \cite{ATLASBs}.
 
Systematic uncertainties are assigned by considering effects not accounted for in the likelihood fit.    The impact of inner detector  residual misalignments was estimated using events simulated with perfect and distorted geometries.   Systematics due to  limitations of the fit model  were determined by  1000 pseudo-experiments  generated with variations in the signal and background mass model, resolution model, background lifetime and background angles models. Systematics due to   $\Bd \ra J/\psi K^{*0}$ and $\Bd \ra J/\psi K\pi$   arise from the  uncertainties of the PDG decay probabilities,  ref.~\cite{PDG}. 

In the absence of initial state flavour tagging the PDF is invariant under the simultaneous transformations: $  \{\phis, \DGs, \delta_{\perp}, \delta_{\parallel}, \delta_S\}  \rightarrow \{    -\phis,  \DGs, \pi - \delta_{\perp}, -\delta_{\parallel}, -\delta_S\}$ leading to a fourfold ambiguity.  As the constraint on $\delta_{\perp}$ is taken from the LHCb measurement~\cite{LHCb:2011aa},  that quotes only two solutions with a positive \phis\ and two \DGs\ values symmetric around zero, two of the four minima fitted in the present non-flavour tagged analysis are excluded from the results presented here. Additionally a solution with negative \DGs\ is excluded following the LHCb measurement~\cite{Aaij:2012eq} which determines the \DGs\ to be positive. The measured values, for the single minimum resulting from these constraints, are given in Table \ref{tab:FitResults}.
\begin{table}[htdp]
\begin{center}
\begin{tabular}{|c|c|c|c||c|c|c|c|}
\hline
Par. & Value & Stat. & Syst.  & Par. & Value & Stat. & Syst.    \\
\hline\hline
 \phis (rad) & \pSfit & \pSstat & \pSsyst &
 $|A_{0}(0)|^2$ & \Azesqfit  & \Azesqstat & \Azesqsyst \\
 \DGs (ps$^{-1}$) & \DGfit & \DGstat & \DGsyst &
  $|A_{\parallel}(0)|^2$ & \Apasqfit & \Apasqstat & \Apasqsyst \\
 \Gs (ps$^{-1}$) & \GSfit & \GSstat & \GSsyst &
$|A_{S}(0)|^2$ & \ASsqfit & \ASsqstat & \ASsqsyst \\
\hline
\end{tabular}
\caption{Fitted values for the physics parameters along with their statistical and systematic uncertainties. \label{tab:FitResults}}
\end{center}
\end{table}
\begin{figure}[htbp]
\begin{center}
\includegraphics[width=0.45\textwidth]{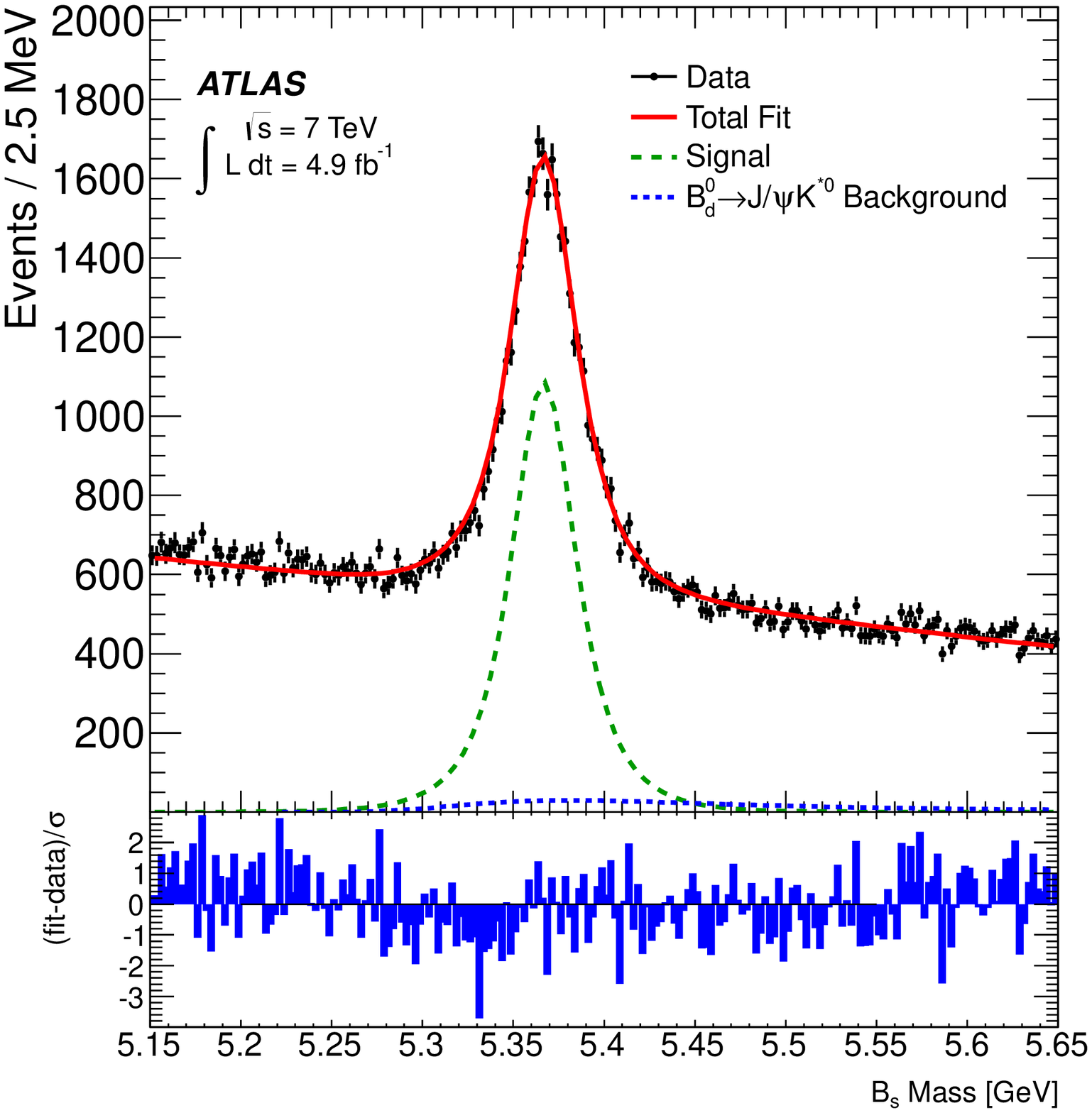}
\includegraphics[width=0.45\textwidth]{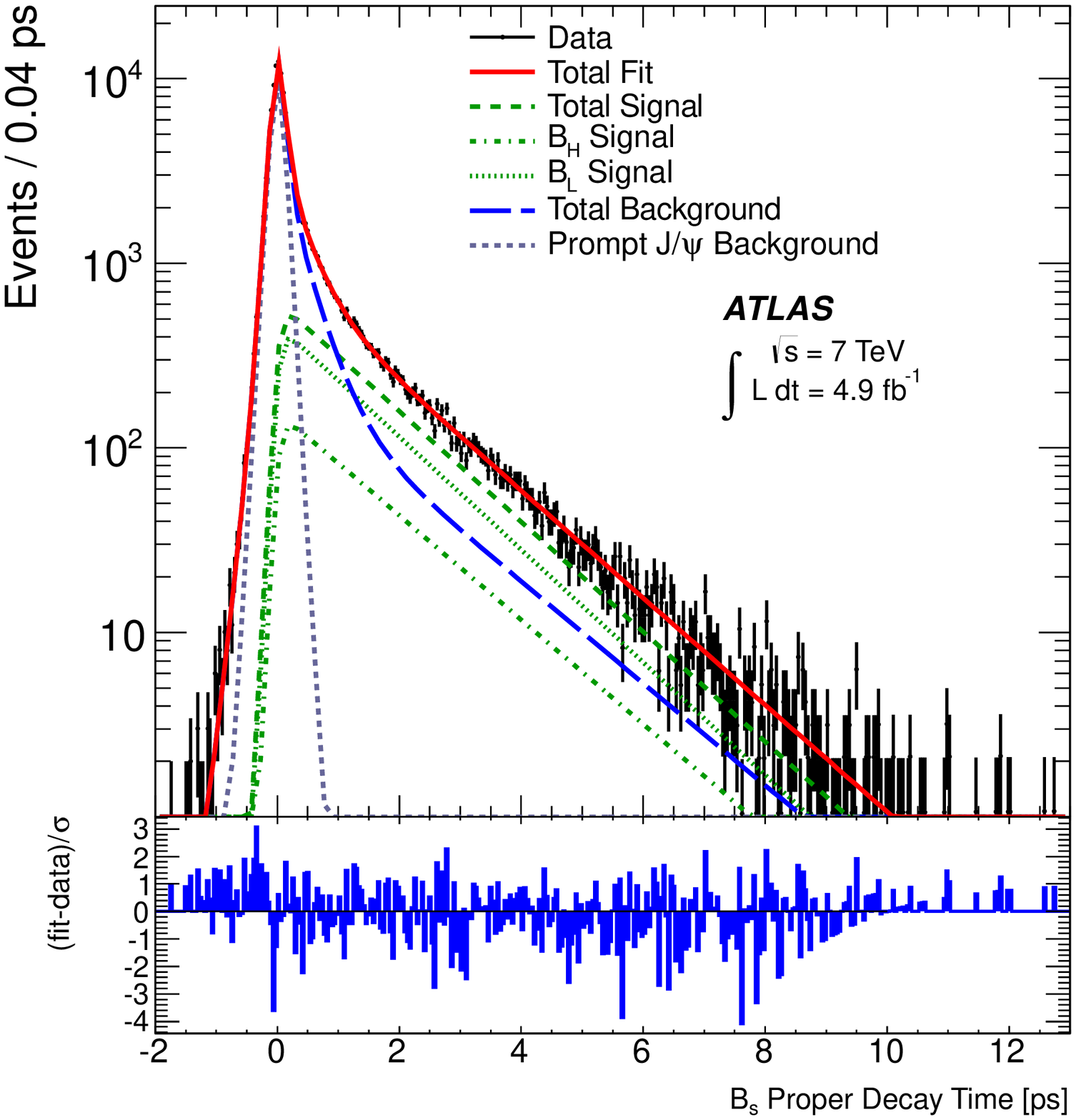}
\caption{ Mass and decay time fit projections for the \Bs \ candidates. The pull distribution at the bottom shows the difference between the data and fit value normalised to the data uncertainty. }
\label{fig:simulFittime}
\end{center}
\end{figure}
The second strong phase, $\delta_{||}$, is fitted very close to its symmetry point at $\pi$. Pull studies, based on pseudo-experiments using  input values determined  from the fit to  data,  return a non-Gaussian pull distribution for this parameter. 
 For this reason the result for  $\delta_{||}$  is  given in the form of a $1\sigma$ confidence interval $[3.04, 3.24]$~rad. The strong phase of the $S$-wave component is fitted relative to $\delta_{\perp}$,  as $\delta_{\perp} - \delta_S = (0.03 \pm 0.13)$~rad.
 The fraction of $S$-wave $KK$ or $f_{0}$ contamination is measured to be consistent with zero, at $|A_S(0)|^2 = \ASsqfit \pm \ASsqstat$. 
The two-dimensional likelihood contours in the \phis\ $-$ \DGs\ plane for the 68\%, 90\% and 95\% confidence intervals are produced using a profile likelihood method and are shown in figure \ref{fig:simulContour}. The systematic errors  are not included in figure \ref{fig:simulContour}  but as seen from table \ref{tab:FitResults} they are small  compared to the  statistical errors. 
The ATLAS measured   parameters of the \Bst\ decay,  using \ilumi\  of integrated luminosity collected in 2011, are  consistent with the world average values and with theoretical expectations, in particular \phis\ is within $1\sigma$ of the expected value in the Standard Model.
\clearpage
\begin{figure}[htbp]
\begin{center}
\includegraphics[width=0.45\textwidth]{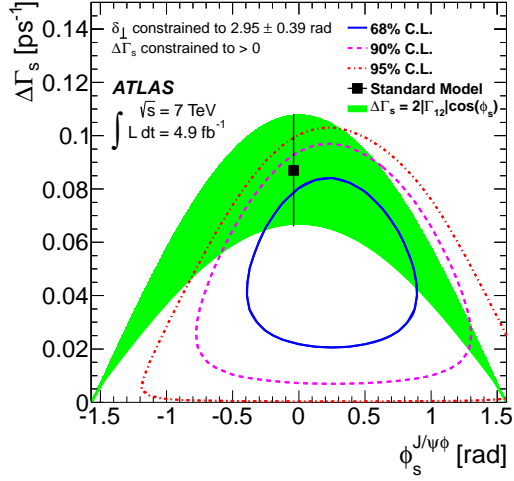}
\caption{ Likelihood contours in the \phis\ $-$ \DGs\ plane, statistical errors only. The green band is the theoretical prediction of mixing-induced $CP$ violation.}
\label{fig:simulContour}
\end{center}
\end{figure}

\bibliographystyle{JHEP}
\bibliography{atlaspaper}
\end{document}

%% file: econfmacros.tex
\def\beq{\begin{equation}}
\def\eeq#1{\label{#1}\end{equation}}
\def\eeqn{\end{equation}}

\def\beqa{\begin{eqnarray}}
\def\eeqa#1{\label{#1}\end{eqnarray}}
\def\eeqan{\end{eqnarray}}

\def\sst{\scriptscriptstyle}

\let\bar=\overbar

\def\Dslash{\not{\hbox{\kern-4pt $D$}}}
\def\dslash{\not{\hbox{\kern-2pt $\del$}}}

\def\msb{{\bar{\ssstyle M \kern -1pt S}}}



%% file: keys.tex
\newcommand{\ilumi} {$ 4.9\, \rm fb^{-1} $}

\newcommand{\DGfit}     { 0.053 }

\newcommand{\DGstat}    { 0.021 }
\newcommand{\DGsyst}    { 0.010 }

\newcommand{\pSfit}     { 0.22 }

\newcommand{\pSstat}    { 0.41 }
\newcommand{\pSsyst}    { 0.10 }

\newcommand{\GSfit}     { 0.677 }

\newcommand{\GSstat}    { 0.007 }
\newcommand{\GSsyst}    { 0.004 }

\newcommand{\Azesqfit}     { 0.528 }

\newcommand{\Azesqstat}    { 0.006 }
\newcommand{\Azesqsyst}    { 0.009 }

\newcommand{\Apasqfit}     { 0.220 }

\newcommand{\Apasqstat}    { 0.008 }
\newcommand{\Apasqsyst}    { 0.007 }

\newcommand{\ASsqfit}      { 0.02 }

\newcommand{\ASsqstat}     { 0.02 }
\newcommand{\ASsqsyst}     { 0.02 }

\newcommand{\cutMassJpsiBB} {$ (2.959-3.229) ~\rm GeV $}
\newcommand{\cutMassJpsiEB} {$ (2.913-3.273) ~\rm GeV $}
\newcommand{\cutMassJpsiEE} {$ (2.852-3.332) ~\rm GeV $}

\newcommand{\cutChiJpsi}   {\chisq $< 10$}

\newcommand{\cutChiBs}   {\chisq $< 3$}

\newcommand{\BspdgNoerr}  {$ 5.3663 ~\rm GeV $}

\newcommand{\Bst}{\ensuremath{ \Bs \ra \JpsiPhi }} 
\newcommand{\Bstbold}{\boldmath{ \Bs \ra \JpsiPhi }} 
 
\newcommand{\Bsto}{\ensuremath{ \Bs \ra J/\psi (\mumu) \phi(K^+K^-) }} 
 
\newcommand{\Bdt}{\ensuremath{ \Bd \ra J/\psi K^{*} }}

\newcommand{\JpsiPhi}{\ensuremath{ J/\psi\phi }}

\newcommand{\Gs}{\ensuremath{ \Gamma_{s}     }}
\newcommand{\DGs}{\ensuremath{ \Delta\Gamma_{s}     }} 

\newcommand{\phis}{\ensuremath{\phi_{s}  }} 
 
\newcommand{\BH}{\ensuremath{B_{H}  }} 
\newcommand{\BL}{\ensuremath{B_{L}  }} 
\newcommand{\GL}{\ensuremath{\Gamma_{L}  }}

\def \ra{\ensuremath{\rightarrow}}

\newcommand{\BsaBs}{\ensuremath{\Bs - \overline{\Bs}}\space}

\newcommand{\aBsn}{\ensuremath{\overline{\Bs}}}

\newcommand{\Jpsipa}{\Jpsi \ }

\newcommand{\chisq}{$\rm \chi^{2}$/d.o.f. }

\newcommand{\Jpsitomm}{\ensuremath{ \Jpsi \ra \mumu }\space}

\newcommand{\phiKK}{\ensuremath{ \phi \ra \kplus \kminus}\space}

